\documentclass[%
superscriptaddress,
nofootinbib,
 amsmath,amssymb,
 aps,
 twocolumn,
prb,
]{revtex4-2}
\usepackage{graphicx}% Include figure files
\usepackage{dcolumn}% Align table columns on decimal point
\usepackage{bm}% bold math
\usepackage{xcolor}
\usepackage{tabularx}
\usepackage{amsmath}
\usepackage{gensymb}
\usepackage{upgreek}
\usepackage{siunitx}
\usepackage{array}
\usepackage{color,soul}

\usepackage{float}
\usepackage{enumitem}
\begin{document}

\title{Intrinsic excitations and a proposed ground state in an Ammann–Beenker artificial spin ice}

\author{E. Weightman}
\affiliation{Department of Physics, University of Liverpool, Liverpool L69 3BX, United Kingdom}

\author{L. O'Brien}
\affiliation{Department of Physics, University of Liverpool, Liverpool L69 3BX, United Kingdom}

\author{S. Coates}
\affiliation{Department of Physics, University of Liverpool, Liverpool L69 3BX, United Kingdom}

\date{\today}

\begin{abstract}
Artificial spin ices built on quasiperiodic geometries remain under-explored. Here we investigate an ASI based on the octagonal Ammann–Beenker tiling, whose six vertex types present local environments inaccessible in periodic lattices. Combining Monte Carlo simulations of dipolar-coupled macrospins with micromagnetic calculations of individual vertex energies, we find that vertex types freeze in an order set not by coordination number, but by quasi-degenerate levels at five-island vertices and their local environment. Excitations in the low energy state are correspondingly localised to specific, geometrically determined sites. Using twice-inflated substitution rules as an energetic backbone, we propose an analytical ground state combining a long-range ordered fixed spin network with a sparse set of degenerate regions.
\end{abstract}

\maketitle

\section{Introduction and motivation}
Artificial spin ices (ASIs) are magnetically frustrated materials \cite{wang2006artificial} which were initially developed to mimic similarly frustrated crystals, such as the rare earth pyrochlore Ho$_2$Ti$_2$O$_7$ \cite{harris1997geometrical}. In general, the alignment of spins in a spin ice is dependent on the geometry of system, where magnetic frustration naturally arises when lattice geometry prevents specific arrangements from occurring. ASIs can possess two types of frustration: vertex, and topological. Vertex frustration arises at sites (vertices) wherein the interactions between individual pairs of spins cannot be simultaneously satisfied,  i.e., if 4 spins meet at a vertex, all pairs cannot be anti-aligned simultaneously. Local energetic environments are then minimised by spins arranging into confugrations with equal numbers `in' and `out', enforcing the so-called ice rules \cite{bernal1933theory}. Topological frustration occurs in geometries which constrain groups of vertices such that each vertex cannot simultaneously exist in their lowest energy state. Vertex frustration is present in most ASI lattices, including the well explored square \cite{wang2006artificial,drisko2017topological, brevis2021topological, porro2013exploring, zhang2013crystallites, ostman2018interaction, kapaklis2012melting} and kagome \cite{li2022geometric, hugli2012artificial, canals2016fragmentation, rougemaille2011artificial, bhat2016magnetization, qi2008direct, moller2009magnetic, gartside2018realization} types, and topological frustration typically arises in more complex geometries \cite{gilbert2014emergent, morrison2013unhappy}. As a result of these types of frustration and mechanisms to relieve it, ASIs have revealed novel and emergent phenomena, including magnetic monopoles \cite{ladak2010direct}, reversal dynamics \cite{brajuskovic2016real, heyderman2013artificial}, and charge screening \cite{gilbert2014emergent, farhan2016thermodynamics}.
 
The prototypical ASI consists of arrays of nanomagnets or magnetic `islands' which are fabricated with lithography and deposition techniques, using magnetic materials such as permalloy (NiFe), cobalt, or nickel. As such, the geometry of both the arrays and the magnetic islands are essentially free parameters, so that ASIs offer a richly diverse playground with which to explore the magnetic properties of a variety of unique systems, and to assess the effects of magnetic frustration in such systems. Simultaneously, the interactions between islands are well characterised across multiple length scales (e.g. as discretised physical objects or as Ising spins), making simulation of ASI magnetic properties both tractable and physically accurate \cite{skjaervo2020advances}.
The range of possible ASI geometries to explore from an \textit{ab initio} standpoint are therefore limitless. Certain periodic structures have been very well explored, and initial studies into disordered ASIs have demonstrated complex reversal behaviour and potential for use in magnonic devices \cite{frotanpour2020magnetization, cote2023direct, frotanpour2021angular}. However, the study of quasiperiodic ASIs has been limited. 

Quasiperiodic geometries possess long range order, but contain no unit cell, placing them on a structural continuum somewhere between pure periodic and disordered systems. Consequently, these structures can support long-range ordered correlations despite, or because of, local spatial heterogeneity which cannot be found in periodic frameworks. The natural quasiperiodic analogue to the commonly explored 2D periodic geometries can be found in tilings -- coverings of the plane using tiles fitted together with no overlaps or gaps. The catalogue of quasiperiodic tilings is extensive, and ever-expanding \cite{tilingencyclopedia}, and thus serves as a rich playground with which to explore ASI properties. Indeed, we note that through the application of local tile rearrangements, even a single base tiling can be systematically tuned to produce entirely new local vertex environments -- effectively expanding one lattice into a rich family of structurally distinct yet related geometries.

Initial work exploring quasiperiodic ASIs has proven intriguing, with topologically induced emergent frustration observed in a Penrose tiling ASI \cite{barrows2019emergent}, and a proposed ground state solution consisting of a fixed skeleton structure and regions of flippable spins \cite{shi2018frustration}.
Likewise, our previous work exploring a quasiperiodic hexagonal ASI found a low energy state that, while topologically frustrated, was minimally degenerate, and with intrinsic excitations \cite{weightman2026artificial}. 
Here, we further broaden the scope of quasiperiodic ASIs by focusing on the octagonal Ammann–Beenker (AB) tiling as a base lattice. Our choice in this instance is essentially arbitrary given the number of possible tilings to use as a base geometry. However, we note that the exploration of the physical properties of the AB tiling continues to unearth characteristic phenomena \cite{singh2024hamiltonian}, and its vertex types offer intriguing local environments compared to previously explored quasiperiodic geometries. Likewise, while a small patch of the AB tiling has already been patterned as an ASI and imaged with XPEEM \cite{bhat2023spin}, there has not been a rigorous analysis of the properties emergent in a thermally-annealed array. 

In this work, we first acquaint ourselves with certain structural properties of the AB tiling before contextualising the geometry in terms of the relevant ASI parameters. Then, we discuss our computational model for exploring the ground state of the AB ASI. Next, we present our numerical results in which we assess the vertex energy and finite charge landscape in order to form the lowest energy configuration of the AB tiling. Finally, we use these results to propose an analytical ground state solution. 

\subsection{The AB tiling as an ASI}

\begin{figure}
	\centering
	\includegraphics[width=0.9\linewidth]{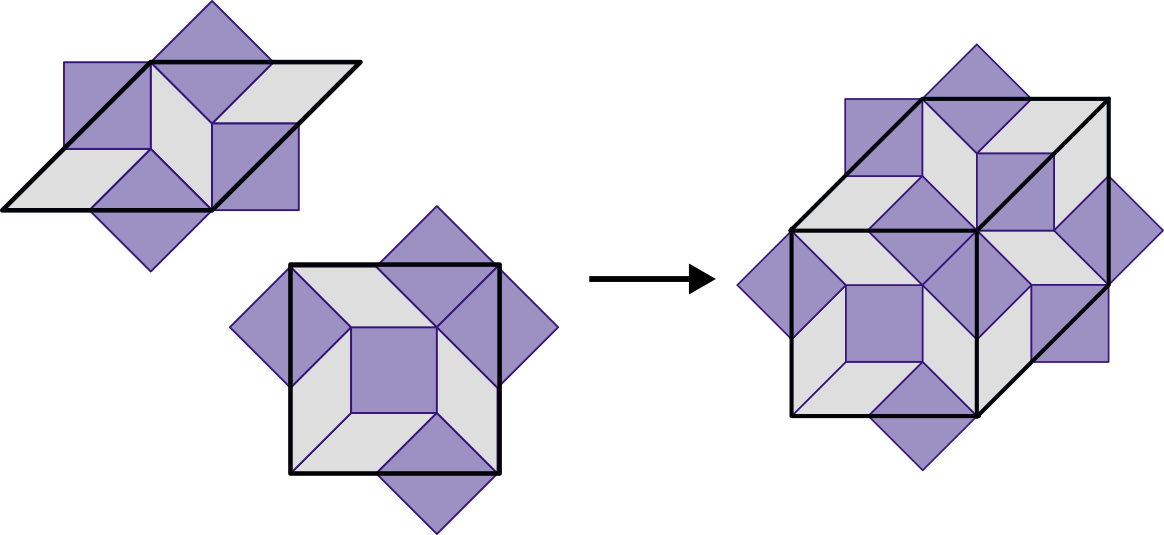}
	\caption{Substitution rules for the AB tiling. The square and rhomb tiles are decorated with deflated versions of themselves which may then be further substituted to form a tiling patch.}
	\label{fig:subtiles}
\end{figure}

\begin{figure*}
	\centering
	\includegraphics[width=0.9\linewidth]{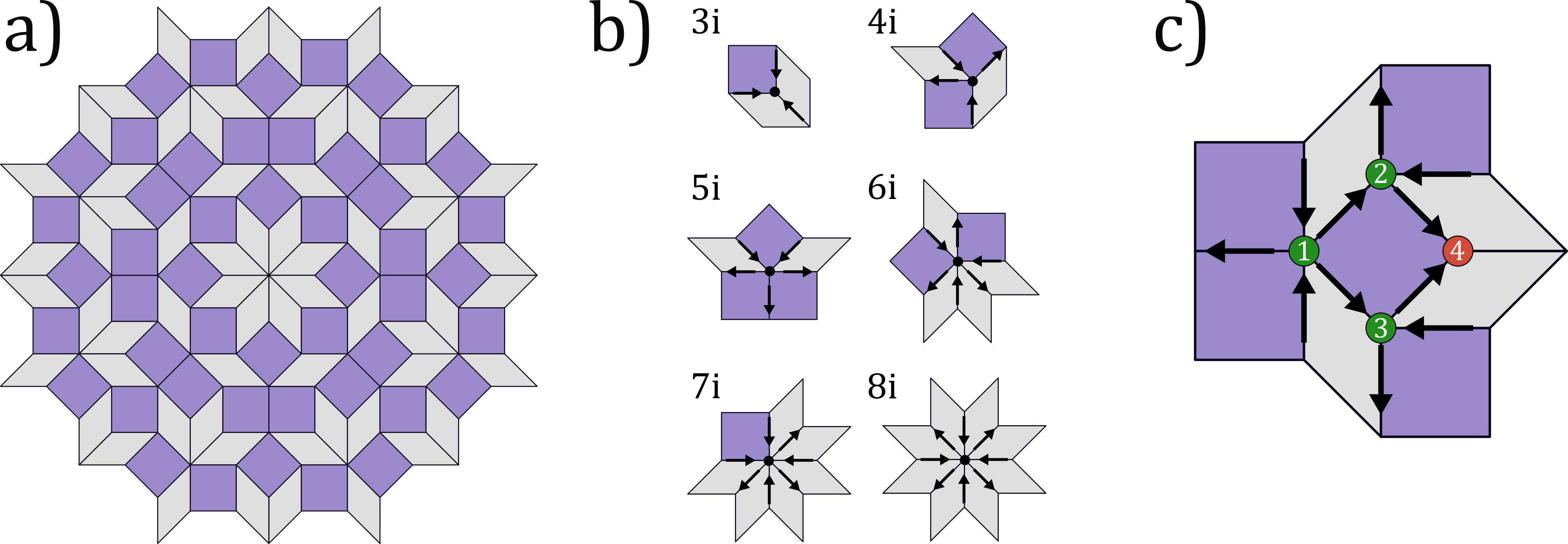}
	\caption{(a) A patch of the quasiperiodic octagonal AB tiling. (b) The six vertex types in the tiling decorated with spins at the midpoints of tile edges and labelled with their coordination number. (c) A patch of the tiling demonstrating topological frustration. The spin configurations which form the lowest energy states for vertices 1, 2, and 3 enforce that vertex 4 must be excited. No arrangement of spins can satisfy all 4 vertices simultaneously.}
	\label{abtile}
\end{figure*}

The AB tiling has 8-fold rotational symmetry and is built from two prototiles that share a single edge length, a square and a 45\textdegree{} rhomb. Patches of the tiling can be constructed in several ways: by the cut-and-project, dual-grid, or substitution methods  \cite{jagannathan2024properties, grunbaum1987tilings}. Here, we briefly recap the latter: Figure \ref{fig:subtiles} shows the substitution rules for the AB tiling. On the left side we see two supertiles in bold black outlines which define the rules -- these are decorated by the square and rhomb prototiles, coloured as purple squares and grey rhombs. The relative scale factor between the edge lengths of the proto- and supertiles is $\lambda = 1+\sqrt{2}$. To build a patch of the tiling, one starts with a prototile (or valid patch of prototiles which belong to the tiling), and replaces each tile with an appropriately scaled supertile copy. An example is given in the right side of Figure \ref{fig:subtiles}, where three tiles (a square and two rhombs) have been replaced by their respective supertiles. Successive substitutions using these rules can then generate larger tiling patches. A small patch of the AB tiling, centred at a position with 8-fold symmetry, is shown in Figure \ref{abtile}(a).

The most natural way to form the AB ASI (or any tiling-based ASI) is to decorate the edges of tiles with magnetic islands arranged parallel to the edge direction, which we discuss in Section \ref{sec:methods}. However, as the bulk of magnetic interaction in ASIs occurs where islands meet tip-to-tip, we note that this choice means that the most relevant structural property of any tiling are the vertex types. In other words, the distinct number of ways in which tiles can meet at a single vertex, as constrained by the substitution rules. Therefore, it is important to contextualise the system in terms of these arrangements, which in turn allows us to make some \emph{a priori} assumptions on the magnetic properties of the ASI array. The six vertex types of the AB tiling with coordination numbers 3-8 are shown in Figure \ref{abtile}(b). The edges of the tiles surrounding each vertex are decorated with arrows which correspond to our magnetic islands. As we are speaking in the language of ASIs, we label each vertex type using its coordination number in the form e.g. `3i', where i refers to nanomagnetic island. The frequency $f$ of each vertex type across the tiling is $\{f_{\text{3i}}, f_{\text{4i}}, ..., f_{\text{8i}}\}$: $\left\{\dfrac{1}{\lambda}, \dfrac{2}{\lambda^2},\dfrac{2}{\lambda^3},\dfrac{2}{\lambda^4},\dfrac{1}{\lambda^5},\dfrac{1}{\lambda^4}\right\}$, such that the 3i is the most prevalent, and the 7i the least.

We can consider the two modes of magnetic frustration (vertex and topological) across an AB ASI array through the lens of these spin arrangements alone, purely from an analytical standpoint. For vertex frustration, we note that trivially, the interactions at a vertex in an ASI are dependent on the geometry. Each vertex in the AB geometry experiences vertex frustation, as numerous islands meet at a vertex -- so, they cannot simultaneously anti-align. The geometry of individual vertices adds additional complexity to the energetics of satisfying the vertices due to their varied coordination, angular separation, and distribution throughout the tiling. Their level of frustration and degeneracy in this regard can broadly be described in terms of the vertex coordination parity. An even number of islands can easily be satisfied by the ice rules, whereas an odd number may not have such a simplistic solution: instead following a pseudo-ice-rule structure, which in turn increases degeneracy and hence entropy. 

From this standpoint, we note that the AB tiling has an equal split in parity: three odd vertices, and three even -- in contrast to the Penrose tiling which has five and two, respectively. In this case, one may naively expect less overall frustration in the AB case. However, of course, the islands decorating an even vertex may not necessarily be isotropically distributed -- while the 8i case consists of eight magnets arranged at multiples of $\frac{2\pi}{8}$, the 6i can be considered as two clusters: one of two spins and another of four, where each spin in a cluster is separated by $\frac{2\pi}{8}$ \footnote{Alternatively, one can view a 6i as an 8i with `missing' spins at the $\frac{2\pi}{8}$ and $\pi$ positions.}. While `solving' the ice rules for these individual clusters is trivial, one may expect a more complex energetic landscape when we consider that interactions are also dependent on the local environment of a vertex i.e., on which vertices these individual clusters are connected to.

This is broadly the definition of topological frustration: additional degrees of frustration arise dependent on the local environments of collections of vertex types. We demonstrate a simple example by studying a common local environment -- Figure \ref{abtile}(c) shows a cluster of 3i, 4i, and 5i vertices which naturally produce topological frustration. In this instance, we have assumed that vertices are satisfied purely according to the ice rules, and have decorated them accordingly. The 5i and 4i vertices are indicated with green circles, and are numbered as 1--3. The 3i vertex is marked in red, as 4. In this example, no decoration of spins may satisfy all vertices  simultaneously, which leads to the excited state. Flipping one of these spins would satisfy the 3i vertex, but in turn excites one of the 4is, and so on. 

\section{Methods}\label{sec:methods}

\subsection{Array-level simulation}

We explore the AB ASI using a Metropolis Monte Carlo algorithm to model a physical system of macroscopic nanoislands. Such a physical system can be realised using lithography to form an array of magnetic islands in a given geometry, and inducing a low energy state through processes such as annealing \cite{gilbert2014emergent, canals2016fragmentation, zhang2013crystallites, drisko2015fepd} or demagnetisation \cite{ke2008energy, nisoli2010effective, morgan2013real, brajuskovic2016real}. 
The physical system we model is formed of permalloy Ni$_{80}$Fe$_{20}$ nanoislands with a saturation magnetisation of $8.6\times10^5$ Am$^{-1}$ \cite{brajuskovic2016real} and dimensions 400nm$\times$100nm$\times$10nm. Our islands are placed at the centre of each tile edge, arranged parallel to the edge direction. The edge lengths of the tiles are then 500 nm, such that the islands are spaced 50nm from the vertices of each tile and do not touch. The islands are considered monodomain, in that each island consists of one macrospin constrained to one of two directions along the axis of the tile edge. The simulated geometry was generated using HyperTiler \cite{hypertiler}, and consists of 1856 tiles, leading to 1857 complete vertices and 3824 spins. Similarly, the system has open boundaries to better mirror a physically fabricated system. 

We simulate 20 separate arrays in the ensemble, where each spin array is initialised with randomised spin directions, with different seeds used for each initial array state. We calculate the local energy of the spin state using a dipolar Hamiltonian comprised of two terms:

\begin{equation}\label{equ:ham}
	\begin{split}
		H = \quad &-J  \; \sum_{\left<ij\right>} \: S_i \cdot S_j \\&+ \: \frac{D}{2} \: \sum_{ij} \left[ \frac{ S_i \cdot S_j}{r_{ij}^3} - \frac{3 \cdot (S_i \cdot \bar r_{ij}) \cdot (S_j \cdot \bar r_{ij})}{r_{ij}^5} \right]
	\end{split}
\end{equation}

Here, $S_{i}$ and $S_{j}$ are the vector spins, with $\hat{r}_{ij}$ and $r_{ij}$ being the vector and distance between each spin pair respectively. We vary coupling across the system, applying the first term to nearest neighbour interactions only as the lattice geometry varies island separation such that a dipolar approximation may break down. As such, we define nearest neighbours as those within a set distance from each spin, which includes only spins whose adjacent angle is $<$90\textdegree{}. This gives a variable nearest neighbour environment across the geometry. All spin interactions include the dipolar term, coupling each spin pair as point dipoles with their centre being the edge midpoint. We truncate the range of dipolar interactions to include only spins within the nearest 5 unique distances, as analysis of the range of dipolar interactions indicated energy contributions beyond this are insignificant. The two terms are governed by the constants J and D respectively, which we calculate for a physical interaction of two islands of the dimensions given, separated by 45\textdegree{}. This gives a ratio of  $\lvert $J/D$\rvert $ = 1.14. Previous studies have found this method reliable for reproducing experimental phenomena  \cite{canals2016fragmentation, rougemaille2011artificial}. 

To reach a low energy state, $10^{6}$ spin-flips are attempted per temperature step through a logarithmic series of 180 temperatures. Spins are selected randomly to flip their alignment, and a flip is accepted if energetically favourable, $\Delta E < 0$, or with probability $P=e^{-\frac{\Delta E}{T}}$. We omit the $k_{B}$ term here as our system is unitary, and we use temperatures scaled by the nearest neighbour coupling constant, J, such that the highest energy interactions are normalised to 1. 

\begin{figure*}
	\centering
	\includegraphics[width=0.9\linewidth]{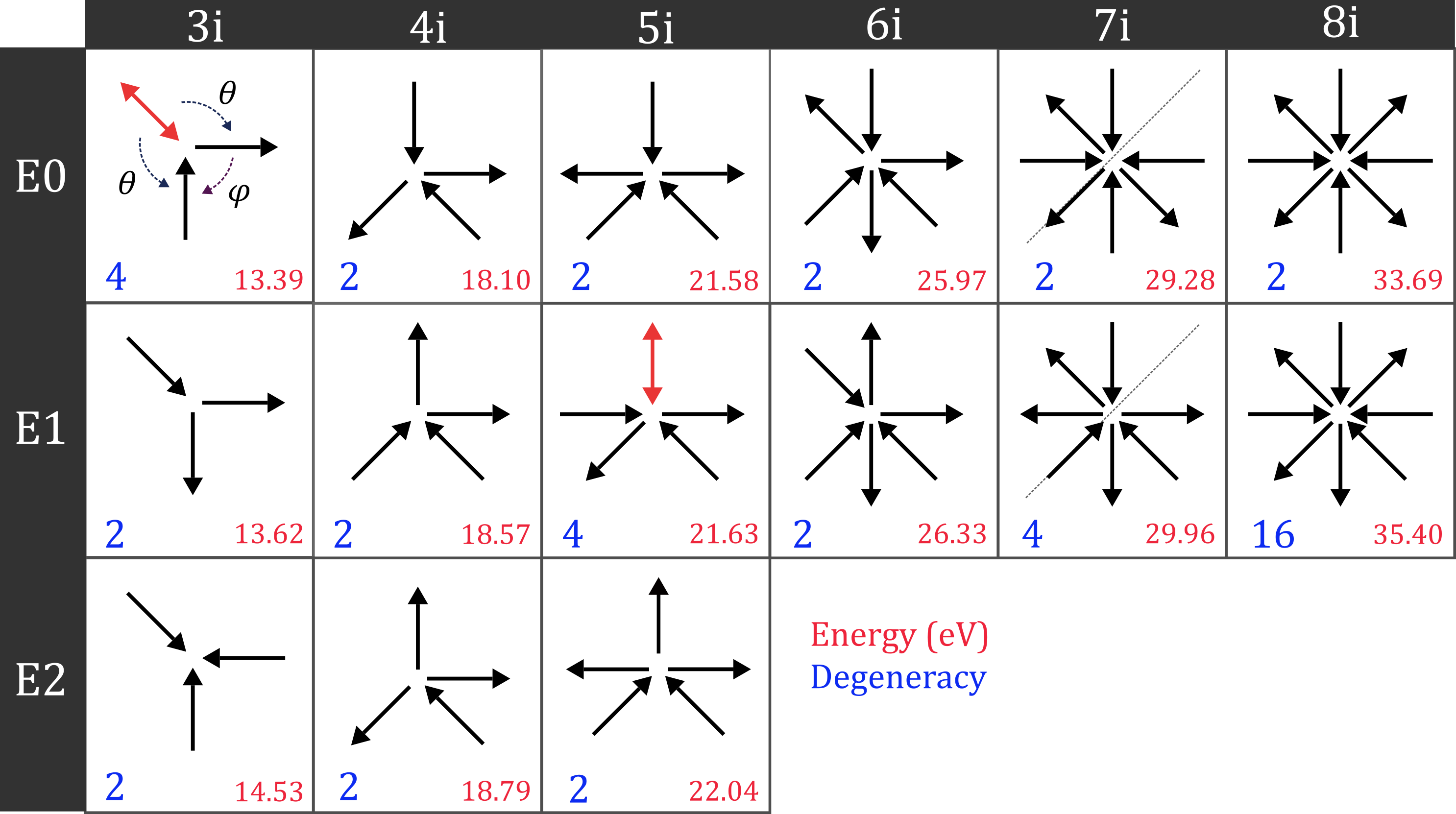}
	\caption{Lowest energy (E0), first excited (E1) and second excited states (E2) for the 6 vertex types in the AB ASI, calculated using the MuMax3 micromagnetic software package \cite{vansteenkiste2014design}. The energy of each configuration is given in red. The degeneracy of each configuration is given in blue, with spins which may point in either direction coloured red.  }
	\label{vertexenergy}
\end{figure*}

\subsection{Vertex-level simulation}
To complement our Monte Carlo simulation, we also perform micromagnetic simulations to characterise the energy of each vertex type only, using mumax3 \cite{vansteenkiste2014design}. This approach individually resolves the interactions between islands physically using a finite difference method, giving the total interaction energy for each configuration. As such, we can analyse the energy states of our simplified point dipole spins with an independent quantitative method. We simulate every possible configuration of spins at each vertex type to identify the lowest energy state, designated E0, the first excited state, E1, and so on. For example, in an 8 spin vertex, we simulate every combination of each spin pointing toward and away from the vertex, with 2$^{8}$ possible states. This may be reduced by half as the states are degenerate if all spins are reversed. The islands are modelled in 2D by imposing constant magnetisation in the z-axis, and are of the same properties as the physical system mentioned previously, ferromagnetic ellipsoidal Permalloy islands of dimensions 400nm$\times$100nm, placed 50nm from the vertex.

\section{Results and Discussion}

Here, we first discuss the energy states of the vertex types in the AB ASI and their features, before using this to assess the vertex energy landscape of the final simulated spin states, and the dynamics of how these states were reached. We then review the average excitation and finite charges on each vertex to characterise how vertices behave spatially, and use this to propose an analytical spin configuration for the AB ASI geometry. We map this to one final spin state, assessing the formation of domains and locations of excitations, and how the long range ordering of spins is propagated in a single domain low energy state.

\subsection{Vertex spin state energies}

Figure \ref{vertexenergy} shows the results of the micromagnetic simulations, with the first two energy states for each vertex type shown. The E2 (second excited) states are also shown for the 3i, 4i, and 5i vertex types. The energy for each configuration is shown in red. The low energy configurations follow the ice rules, in that spins preferentially follow an in-out alternating pattern \cite{bramwell2001spin}. Where this cannot be obeyed due to an odd number of islands, the lowest energy state occurs when the islands with the largest angular separation form an in-in or out-out pair. This is demonstrated in Figure \ref{vertexenergy} for the E0 3i vertex, in which the angle $\theta$ represents the largest angular separation between spins, and the angle $\phi$ the smallest.  The in-in or out-out pair are therefore formed with the indicated red double-arrow (a freely flippable spin with no energy cost) and either of the two other spins, i.e. between $\theta$ separated pairs. Vice versa, the anti-aligned pair occur between $\phi$ separated spins. The 5i vertex is notable in this regard: the strong interaction of the $\phi$=45\textdegree spins creating in-out pairs means an in-in or out-out pair must be assigned to one of the three sets of spins separated by $\theta$=90\textdegree{}. This results in two energy levels which are quasi-degenerate ($\Delta$E = 0.05eV) -- for E0 the in-in pair resides on the lower square tile edges, while for E1 a freely flippable spin creates either an in-in or out-out pair for the two adjacent square tiles.

The degeneracy (number of possible energy states) of each configuration is shown in blue. Each vertex is at least twofold degenerate. However, the 3i E0 and 5i E1 states have a higher degeneracy due to a freely flippable spin, indicated by the red arrows, which may point in either direction for the same energy state. Similarly, the 7i E1 has a degeneracy of 4, due to the same-direction-pair which may lie on either side of the line of mirror symmetry, indicated by a dotted line. The increased degeneracy in the E1 8i vertex is due to its high symmetry, wherein any 4 adjacent spins could form the in-in out-out pairs as they all possess the same angular separation.

\subsection{Vertex energy state populations and dynamics}

\begin{figure}
	\centering
	\includegraphics[width=0.8\linewidth]{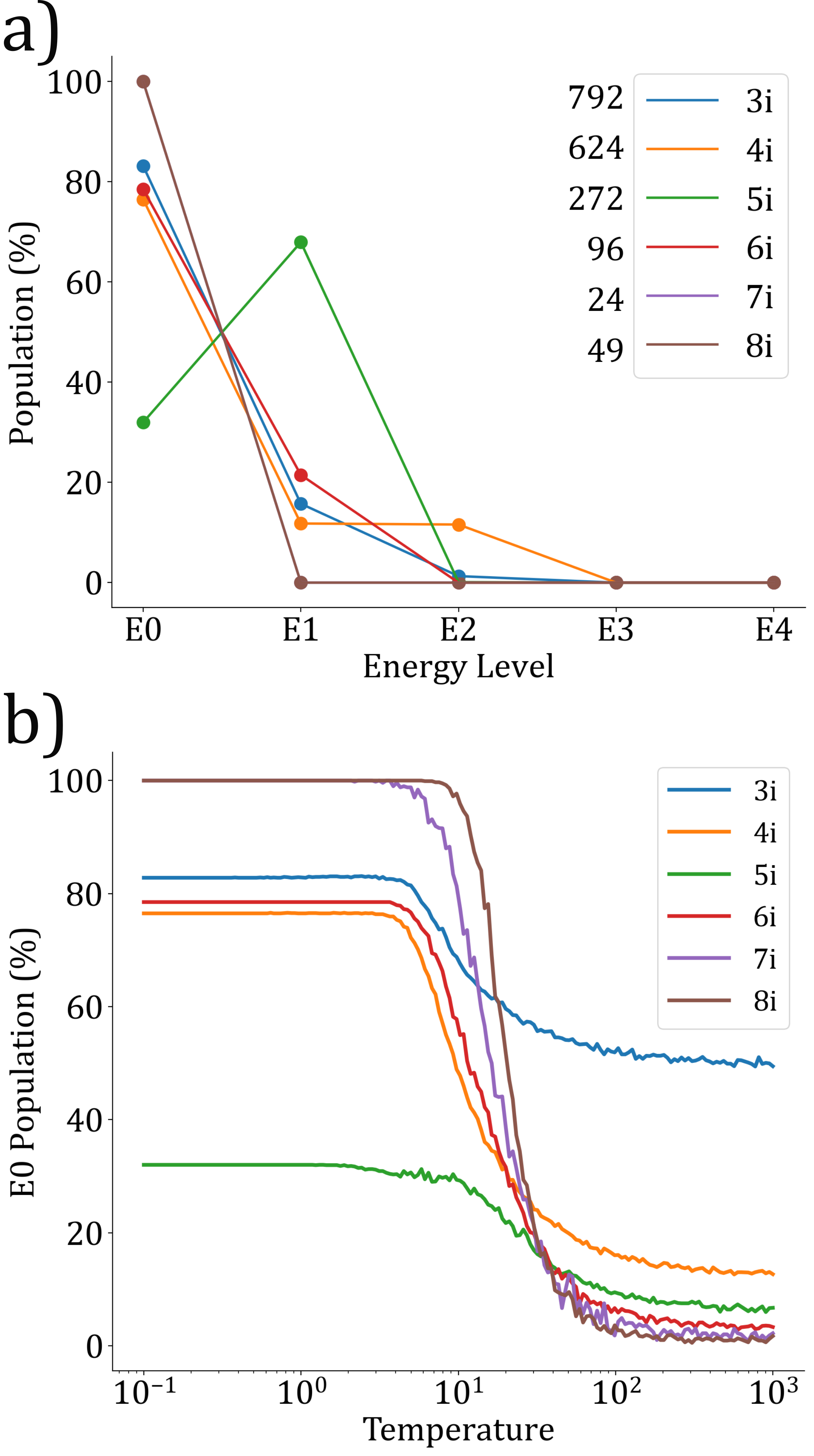}
	\caption{(a) Average energy state populations for each vertex type, denoted by their coordination numbers, with E0 being the lowest energy level. The total number of each vertex type in the simulated patch is shown to the left of the legend. (b) Average E0 population for each vertex type as the temperature is reduced. Vertex types are denoted by their coordination numbers, with each population beginning at a level of  the proportion of E0 configurations in all available states, before increasing as magnetic order begins and finally remaining unchanged in the final configuration.}
	\label{vstats}
\end{figure}

Figure \ref{vstats}(a) shows the average final energy level populations for each vertex type following the Monte Carlo simulation. There are no excitations in the final spin states beyond E2, which is observed for a material number of 4i vertices, and some 3i vertices. E0 is the most populous energy level for all vertex types, except for 5i, which we discuss further on. Similarly, the number of E1 and E2 excitations for the 4i are split evenly, $\sim16\%$ each, which we discuss in the next section. 

To broadly assess how these energy populations are reached, Figure \ref{vstats}(b) shows the average E0 population for each vertex type as the temperature of the system is reduced. The initial vertex configurations are randomised, thus the initial E0 population is determined by the number of possible spin configurations at each vertex. As an example, the 3i vertex has $2^{3}$ possible states, of which 4 are E0, thus the initial E0 population is approximately 50\%. All vertex types show an increase in E0 population from their probabilistic baseline as the temperature decreases, then reach a plateau. The 8i vertices reach a population of 100\% E0 states at a temperature T$_{plat}$, at which point they lock-in and remain unchanged due to the large amount of energy required to excite to E1, Figure \ref{vertexenergy}.

\begin{figure*}
	\centering
	\includegraphics[width=0.99\linewidth]{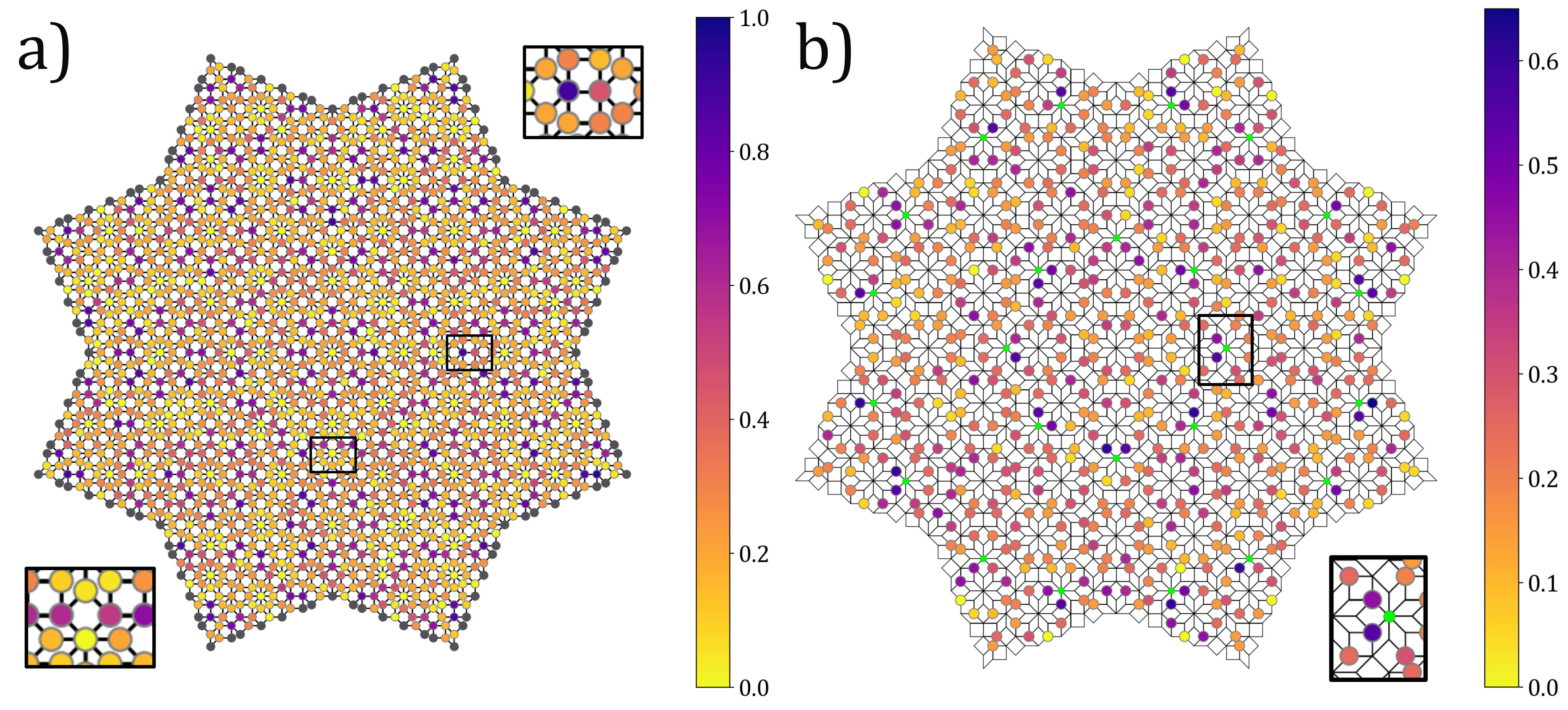}
	\caption{(a) Average energy state for each vertex, calculated with a score of 0 for E0 configurations and 1 for any excited states. Incomplete vertices on the edge are given no score and coloured grey. A darker colour indicates a vertex which is more frequently excited, with yellows, such as those of 8-island vertices, always being E0. (b) Average energy states as in (a) for 4i vertices only. The colour scale is normalised to the maximum 4i score, not 1 as in (a). 7i vertices are highlighted in green.}
	\label{heatmapavEnergy}
\end{figure*}

Aside from the 8i, the 7i, 6i, 4i, and 3i vertices all follow similar behaviour to one another, with the same overall trend in population increase, and a population plateau at similar temperatures, $\sim$T$_{plat}$/3. 
Comparatively, the 5i vertices' E0 population increase is much shallower, and the temperature at which the population ceases to change is also lower than for the other vertex types, at $\sim$T$_{plat}$/4. This behaviour, and the high percentage of E1 states observed, can be directly related to the difference between the E0 and E1 energy levels of the 5i vertex, and its local environment. 
First, this difference is 0.05eV, marking the two levels as nearly degenerate, such that the excitation can easily be accessed at lower T -- which would in turn contribute to the shallower slope. Second, the difference between the E0 and E2 levels for the 5i, 0.46eV, is smaller than the difference between the E0 and E1 levels for the 4i, 0.69eV. The local environment of a 5i is such that at least two spins are connected to two adjacent 4i vertices \cite{jagannathan2024properties}, so it follows that both local and global energy will \emph{decrease} even if a 5i is doubly excited in order to de-excite a neighbouring 4i. Therefore, it is logical that the 5i energy level population continues to vary until the 4i vertices lock-in. Indeed, even if a 5i vertex exists in the E0 state, it is likely to re-excite to satisfy its surroundings. We also note that the freely flippable spin of the 5i E1 state (Figure \ref{vertexenergy}) is a spin which is shared only with adjacent 5i vertex types -- in other words, the 5i vertices exclusively exist in connected pairs along this spin \cite{jagannathan2024properties}. As such, if both 5i vertices are locked into E1 state this spin may reverse continually, even when all other spins are fixed at low T.

The behaviour of 5i vertices means that the ordering of vertex types as the temperature decreases does not follow with coordination number, behaviour which one may naively assume \emph{a priori} given that increased coordination number typically leads to stronger local coupling. Instead, we see an ordering of the form $8i \rightarrow 7i \rightarrow 6i \rightarrow 4i \rightarrow 3i \rightarrow 5i$. This serves to emphasise that the local environment of a single vertex is crucial in how this hierarchy is propagated, and provides insight into how and where excitations may form.

\subsection{Specific vertex state environment behaviour}

While we have focussed on the average states of each type of vertex, we can also assess the behaviour of each specific vertex given their individual local environments. This provides a view of how excitations and charges are propagated given the geometry, as well as indicating if excitations may be intrinsic and present in a lowest energy solution, or, as a result of colliding domains and frozen-in defects which we may expect to disappear in the average. Figure  \ref{heatmapavEnergy}(a) shows the average excited/not excited state of each vertex. Here, we count an excited vertex with a score of 1 and an E0 vertex with a score of 0, computing the average score for each vertex in the ensemble. Darker colours indicate if a given vertex is more frequently excited, and vice versa. Vertices lying on the edge (which are as such incomplete) are not counted as either state.
8i and 7i vertices exist solely in the E0 state and so only score 0, as indicated by their yellow colour in Figure  \ref{heatmapavEnergy}(a). The 3i and 6i vertices show some variation in colour, but there is no specific location or local environment which is consistently more or less excited. The 5i islands appear routinely excited, as discussed, producing a distinct coronal-like excitation map purely based on their positions in the tiling. 

The 4i vertices also contribute to this spatial excitation map, although we find that their behaviour is split, dependent on their local environments. To demonstrate, the isolated excitation scores of 4i vertices are shown in Figure  \ref{heatmapavEnergy}(b), with colour scaled relative to the scores of 4i vertices rather than the system as a whole. The average score of all 4i vertices is 0.23, with a minimum and maximum score of 0 and 0.65 respectively. Qualitatively, we note that a subset of these vertices appear more darkly coloured. We find that this subset consists exclusively of 4i vertices which neighbour a 7i vertex, highlighted in Figure  \ref{heatmapavEnergy}(b) as an inset. Considering vertices exclusively in this location gives an average excitation score of 0.48 -- over double the average for all 4i -- suggesting that this local geometry reliably experiences excitation. Simultaneously, it suggests that both of these sites are reliably excited exactly half of the time, or, that one is excited and another is not in every array, with the choice being probabilistic. Further analysis shows that it is the latter case, with the level of excitation split between E1 and E2; similarly, these positions explain the high percentage of E1 and E2 states observed and discussed in Figure \ref{vstats}(a). Indeed, these observations and their origin highlight the environment around the 7i vertices as a specific region of interest, which we discuss in the next section.

\begin{figure*}
	\centering
	\includegraphics[width=0.9\linewidth]{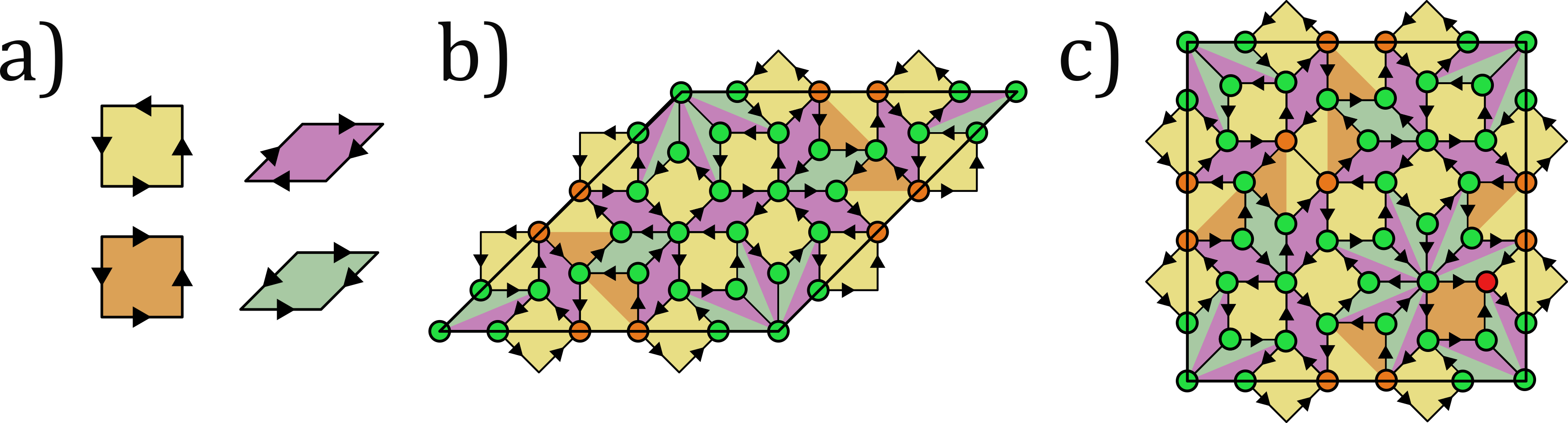}
	\caption{Tiles in the proposed ground state solution. (a) The prototiles which build the proposed solution in one domain. (b) A rhomb supertile decorated with the prototiles in (a). All vertices are in the E0 states (green) except for the 5i on the edges, which are excited to E1 (orange). (c) Square supertile decorated with the prototiles in (a). The majority of vertices are E0 (green) expect for the 5i vertices, excited to E1 as in (b). The region adjacent to the 7i vertex is particularly frustrated, with a 4i excited to E2, shown in red.}
	\label{analyticaltiles}
\end{figure*}

\subsection{Proposed analytical ground-state solution}
Informed by the energy and finite charge landscape of the simulated spin states, we propose a ground state spin configuration for the AB ASI with a set of decorated prototiles and supertiles. The key ingredients from our analysis are that 8i and 7i vertices must be in the E0 states, and that 5i vertices and 4is (connected to 7is), are easily excitable. Considering this, Figure \ref{analyticaltiles}(a) shows four prototiles -- two squares and two rhombs -- which are decorated with arrows indicating spins which occupy one of two degenerate magnetic states. The yellow square, and pink rhombs are decorated with spins that form an anti-, and clockwise sequence, respectively. The green rhomb has two pairs of spins which are anti-clockwise and clockwise, while the orange square has three spins in an anti-clockwise sequence. The opposing magnetic state simply flips each of these spins. 
	
Using these prototiles, we can build the supertiles shown in Figure \ref{analyticaltiles}(b, c), which are used to propagate our proposed solution across an infinite tiling. The internal skeleton of these supertiles are created by simply twice inflating the base AB prototiles, which allows us to decorate each vertex type at least once internally, with the exception of the 8i, which we discuss further on. The half-shaded tiles (pink/green rhombs, yellow/orange squares) indicate positions in which either colour tile can be placed without an increase in energy to the system. The vast majority of vertices sit in their E0 state inside the tiles, with 5-island vertices excited to E1 on the edges of the rhomb, and internally within the square. Attempting to further satisfy the pairs of connected 5i islands on the edges of all supertiles comes at the comparatively high energetic cost of exciting some 3is to E1 and E2 levels, which would add $\sim$2 eV per supertile. The immediate region around each 7i is particularly frustrated, with a 4i vertex excited to E2, as shown by the red circle. This induces asymmetry about the 7i vertex as two 4i are connected to each 7i, but only one is excited. The orientation of the 7i has no impact energetically, such that it may be in either of its E0 states, however, the location of the 4i excitation is determined by this orientation. An arbitrary one of these states is shown in Figure \ref{analyticaltiles}(c), where flipping all spins in the 7i vertex moves the E2 excitation to the opposite side.

It is possible, from a vertex-only standpoint, to decorate the square supertile in Figure  \ref{analyticaltiles}(c) with a lower vertex-by-vertex energy configuration -- exciting and de-exciting several vertices in the region surrounding the 7i. If we purely sum the energy of individual vertices, then exciting a 3i to E1, de-exciting the E2 4i to E1, and de-exciting a 5i to E0 accounts for an overall energy reduction of 0.041eV. However, this decoration is lower only in a system which does not account for dipolar coupling or far field effects, in other words, calculating just a sum of the energy of each vertex configuration independently. In our approach to a ground state solution, we have considered the system under the Hamiltonian we are modelling, such that we propose a more system-specific solution, accounting for the range of interactions we simulate. The vertex-only approach is unphysical when considering a system where neighbouring, but not connected, spins can interact, e.g.: spins on opposite sides of a square tile. 

Appendix \ref{app:square supertiles} shows two additional configurations for the square supertile which are energetically lower from a vertex-only viewpoint, but higher when long range interactions are included. Because of these interactions, it becomes energetically favourable to excite the 4i vertex to the relatively high E2 level, as seen in Figure \ref{analyticaltiles}(c), and demonstrated in Figure \ref{fig:hamil_energy}. This decoration, notably, maximises closed flux loops, which are inherent in the yellow and pink prototiles; this maximises the anti-alignment of spins which lie on opposite sides of each square prototile. In the case of the connected E1 5i vertices, the shared spin is therefore always aligned with one parallel neighbour and anti-aligned with the other.

Although the 8is are not explicitly included within the supertile configurations in Figure \ref{analyticaltiles}(b,c), they are formed wherever the supertiles meet vertex to vertex. We have not decorated the prototiles at the supertile vertices with spins, as the orientation of the 8i vertices does not determine the orientation/domain of the surrounding vertices -- the 8i vertex can essentially exist at its E0 level in either domain without impacting the states of the surrounding tiles. However, this does not mean that the orientation of the 8i vertices has no influence in the simulated ASI, especially given that their state is frozen in first. Rather, their orientation in general, and with respect to one another, does not determine the domain state of the surrounding environment. 

\begin{figure*}
	\centering
	\includegraphics[width=0.9\linewidth]{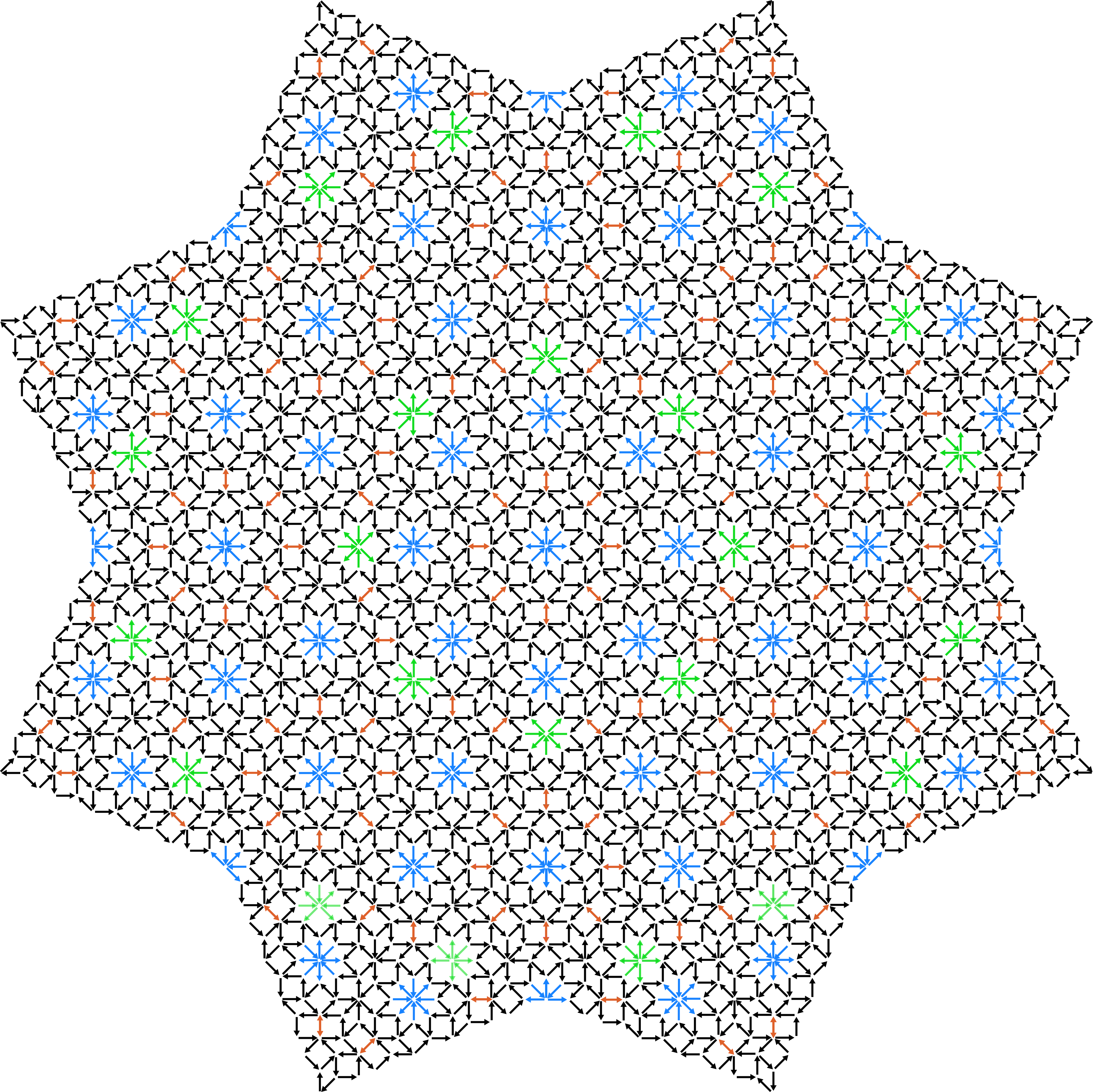}
	\caption{Spins in the proposed ground state solution. Black spins form a fixed network of two fold degeneracy. Blue spins show 8i vertices which must be E0 but whose direction as a group 8 is independent of each other and the fixed network. Orange arrows show freely flippable 5i spins which may point in either direction at no energy cost. Green arrows show spins which form a 7i vertex. Like the 8i, they may exist in either E0 state, determining the location of the E2 4i excitation.}
	\label{fig:sflip}
\end{figure*}

To broadly assess the validity of the analytical solution we compare energies of our simulated data. The total energy difference between the average final simulated state and the analytical array as a sum of vertex energies across the array shows that the proposed solution is lower by ${\sim}108$~eV, or ${\sim}0.06$~eV
per site. This difference can also be contextualised directly in terms of the extrinsic excitations in the simulated states, i.e. those not present in the analytical solution. Writing $N^j$ for the number of type-$j$ vertices, $\Delta f^j$ for the difference between the simulated and analytical frequencies of excited states, and $\Delta E^j = E_1^j - E_0^j$ for the excitation energy where we simply assume first level excitations, the total is
\begin{equation}
	E_\mathrm{tot} = \sum_{j=3}^{6} N^j \, \Delta f^j \, \Delta E^j = 87~\mathrm{eV},
	\label{eq:extrinsic-energy}
\end{equation}
the shortfall presumably arising from our approximation that all excitations are to $E_1$.

Overall, the proposed analytical solution to the ground state of the AB ASI explains the two perhaps unexpected features observed in our simulated systems (significant proportion of excited 5i sites and high proportion of 4i excitations), while faithfully representing the high proportion of low energy states of other vertices. As such, similar to the analytical solution proposed for the Penrose tiling \cite{shi2018frustration}, our ground state exhibits built-in topological frustration with excitations enforced by local geometry, occurring at all 5i vertices and $\sim 5\%$ of 4is. Also similar to the Penrose solution is the feature of long range order in a fixed network of spins, with a collection of `flippable' spins outside of this. In other words, in our solution for the AB geometry the majority of spins (the fixed network) must be of a specific fixed direction to form the lowest energy state. Fixing the orientation of any of these spins thus determines the orientation of all of them. Outside of this fixed long range ordered network are spins which may freely flip; those lying between two E1 5i vertices, and islands forming 7i and 8i vertices, which must exist in the E0 state but may be of either domain. Figure \ref{fig:sflip} shows the simulated array in one of the proposed ground state configurations with these fixed network spins coloured black, and the variable spins coloured orange, green, and blue for the 5i, 7i, and 8i spins respectively.

\begin{figure}
	\centering
	\includegraphics[width=0.9\linewidth]{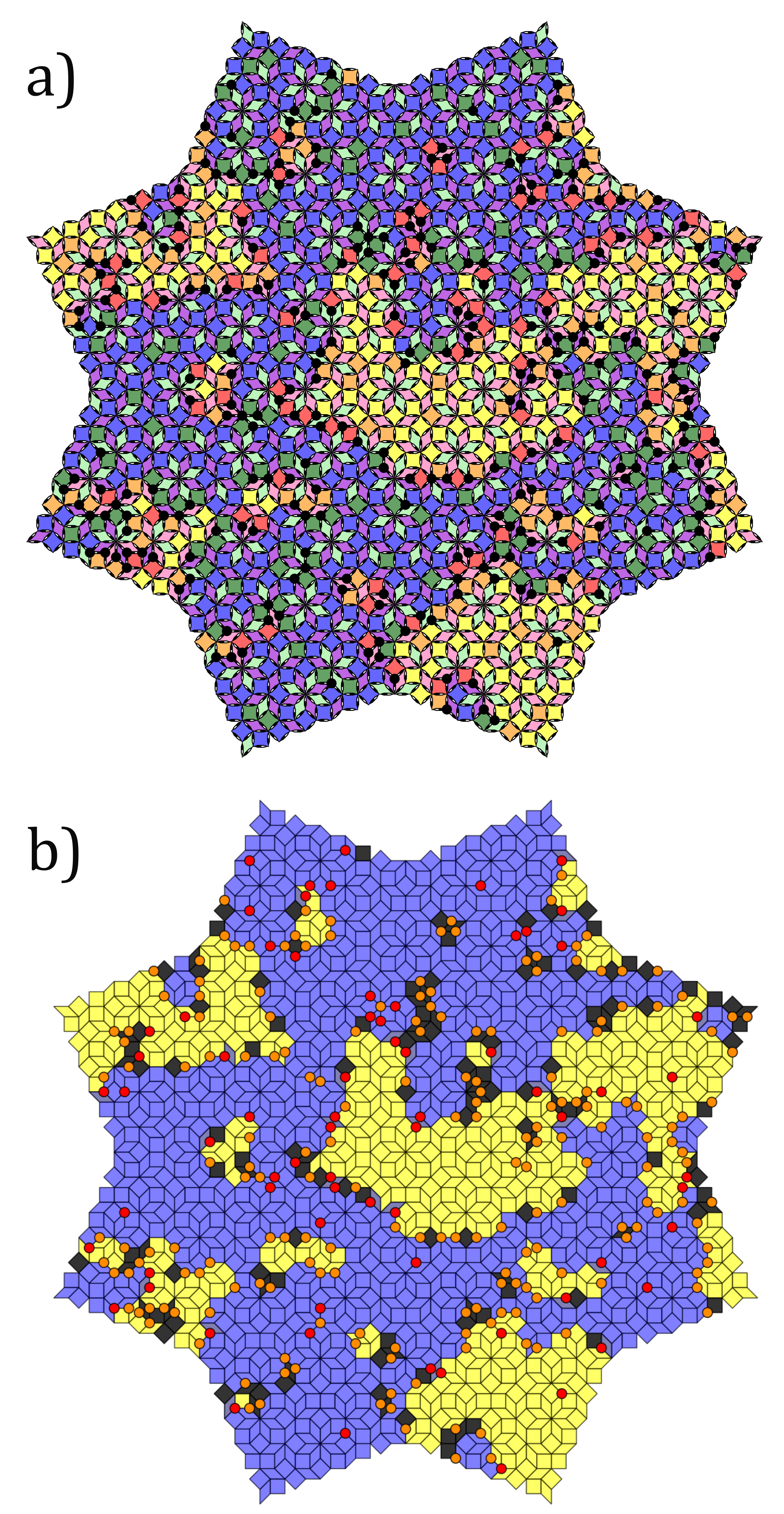}
	\caption{(a) Tiles of one final simulated spin state coloured according to the prototiles in the proposed analytical solution, and the opposite domain. Black circles show excitations. Excitations for 5-island vertices are not shown. (b) Tiles coloured according to the domain with which they match. Excitations are highlighted with orange circles for E1 and red for E2. Excitations for 5-island vertices are not highlighted due to their being built in the analytical solution. Tiles which do not form a domain cluster of three or more are coloured black. }
	\label{domains}
\end{figure}
\subsection{Domain formation using the ground state solution}
%Map this to simulation results
We next compare our proposed analytical solution to one final spin state in order to analyse opposing domains and their boundaries. We define one domain by the orientation of spins in the prototiles in Figure \ref{analyticaltiles}(a), and the opposing domain being formed by flipping every spin. Excitations which are intrinsic to the system would be expected to exist within clusters of same-domain prototiles. Other excitations, which are the result of colliding domains and frozen-in defects, would then be expected to exist on the edge of these clusters.

\begin{figure}
	\centering
	\includegraphics[width=0.9\linewidth]{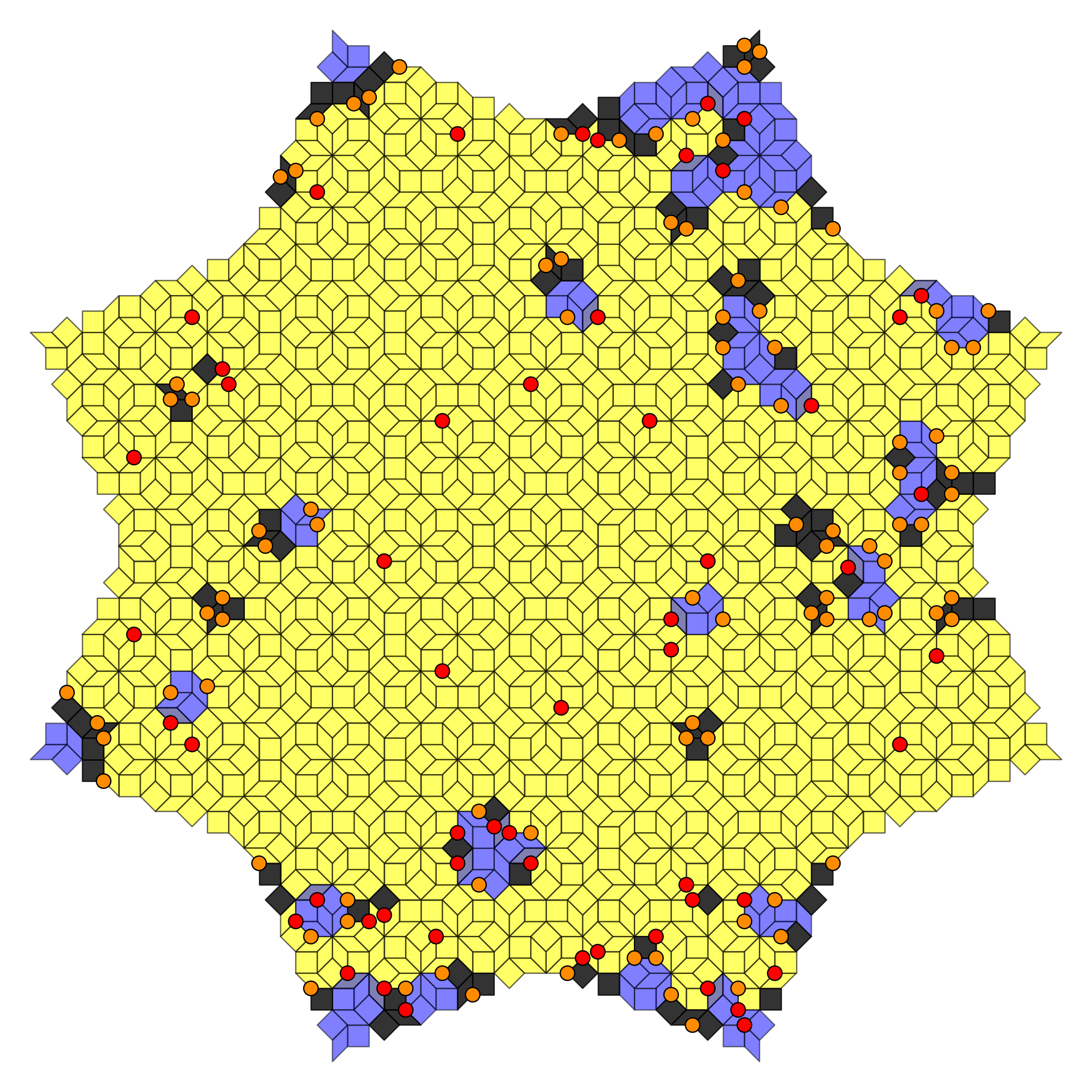}
	\caption{The final state for one simulated system with 6i vertices fixed in one E0 states. Tiles are coloured according to the domains defined by tiles previously, with E1 and E2 excitation coloured orange and red respectively. E1 excitations of 5i vertices are not coloured due to their frequency. Tiles which do not form a domain patch of 3 tiles of more are coloured black.}
	\label{fig:fixed6domain}
\end{figure}
Figure \ref{domains}(a) shows such a mapping, with tiles coloured in the same way as in Figure \ref{analyticaltiles}. Four additional coloured tiles are included here, a blue square tile ( the opposite domain counterpart of the yellow squares -- a clockwise closed loop); a green square tile(the counterpart to the orange square); a purple rhomb (counterpart to the pink rhomb); and a red square (similar to the green rhomb with two clockwise and two anti-clockise pairs of spins). Excited vertices excluding E1 5i excitations are shown in black. Figure \ref{domains}(b) shows the same final state with the tiles coloured according to their overall domain (as defined by the tiles in Figure \ref{analyticaltiles}), with the yellow domain being of the direction shown in Figure \ref{analyticaltiles}, and the blue domain being the equal and opposite. Excitations here (excluding E1 5i) are shown in orange for E1 and red for E2. Tiles which do not fit within a domain patch of three or more tiles are coloured black. By this mapping, we observe the formation of two opposing spin domains with tiles that match those in our proposed analytical solution.  Where the two domains meet, strings and clusters of excitations are seen, forming a domain wall of increased energy and charge -- as expected when two opposing domains meet in similar magnetic studies \cite{budrikis2012domain, skjaervo2020advances}. There are also, in some cases, defects in which clusters of tiles fit to neither domain. As there appears to be no spatial dependency on the appearance of the defect clusters, it presumably occurs when a small number of spins of one domain are trapped by a local environment of the other at a sufficiently low enough temperature and are frozen out.

However, the state in Figure \ref{domains} shows several 6i excitations which do not occur in the analytical solution, but \emph{do} exist purely within clusters of excitations and patches of tiles in neither domain. As such, we assume that these sites are the most easily accessed by colliding domains or frozen--in defects.  Given the fixed spin network in the proposed analytical solution, fixing one such spin should enforce a single domain over the whole simulated structure. As such, we performed additional simulations in which we fixed all 6i vertices into one domain orientation as dictated by the analytical solution. We select the 6i vertices to be fixed as they are the highest coordination vertices which can influence the fixed spin network, as both the 8i and 7i vertices can exist in either E0 state. Likewise, by removing the route to relaxation of colliding domains, we assume we can produce a monodomain.

Figure \ref{fig:fixed6domain} shows the final spin state for one simulated system, with tiles coloured according to the tile domains as in Figure \ref{domains}(b). The majority of tiles are of one domain, coloured yellow, which is the domain enforced by the field on the 6i vertices. Some patches of the opposing state are found, in blue, with corresponding excitations along these boundaries, where orange circles indicate E1 states, and red are E2. Therefore, the fixing of the 6i vertices enforces the formation of a single domain in the \emph{fixed} network, with some isolated defects of the reverse domain still present, particularly on the edge. These defects are not spatially dependent, suggesting they are frozen in stochastically. 

\section{Conclusion}

We have simulated an artificial spin ice composed of monodomain nanoislands using the AB tiling for the lattice geometry. Using Metropolis Monte Carlo simulations and complementary micromagnetic simulations, we found a low energy state which hosts intrinsic excitations at specific vertex sites. Then, using twice-inflated substitution rules as an energetic backbone, we proposed an analytical ground state for an interacting dipolar system which encompasses these excitations while remaining in the energetic minimum. In this manner, our work tallies with previous work on the Penrose tiling \cite{barrows2019emergent}.

The rich physical behaviour we have explored is specific to the AB tiling at both local and environmental levels. The dynamic freezing of the vertex types did not follow coordination number, arising instead from the quasi-degenerate energy levels at the 5i site and from a local environment which uses the 5i as a mechanism to de-excite its neighbours. This follows from the manner in which the substitution rules constrain which vertex types may neighbour one another. Frustration in the AB ASI is therefore not distributed by vertex parity or coordination alone, but is localised by the inflation hierarchy onto a sparse and deterministic set of sites.

This localisation has a direct consequence for magnetic ordering. Because the fixed spin network we propose is rigid, fixing any one of its spins fixes all of them. In practice, however, pinning the 6i vertices propagated a single domain only locally, with reversed clusters frozen-in stochastically in regions distant from the pinned sites. A monodomain state is thus accessible in principle, but is limited by the range of the dipolar interaction relative to the spacing of high-coordination vertices -- either the interaction must be extended, or the density of such vertices increased through phason flips or modified substitution rules.

The predictions we make are directly testable. A small patch of the AB tiling has already been patterned in permalloy and imaged with XPEEM \cite{bhat2023spin}, albeit in the field-driven reversal regime rather than the thermally annealed one, and the annealing protocols established for square and kagome ASI should transfer without modification. Our excitation map provides an unambiguous signature: 5i vertices should appear excited even in a well-annealed array, and the pairs of 4i vertices adjacent to each 7i should be excited exactly half of the time. It is also notable that a $J_1-J_2$ Ising model on the same tiling likewise develops domains pinned to particular quasiperiodic sites \cite{teixeira2025stripe}, suggesting that the tiling, rather than the Hamiltonian, selects where disorder accumulates. In the same spirit, the constrained coverings studied on AB tilings -- dimer coverings and Hamiltonian cycles \cite{singh2024hamiltonian} -- offer a natural combinatorial framework for counting the ice-rule manifold, which we have here approached configuration by configuration.

More broadly, the AB ASI shows that a quasiperiodic spin ice need not trade long-range order against frustration. The inflation hierarchy supplies both at once: a macroscopically ordered fixed network, alongside a countable and geometrically determined set of frustrated sites whose density is set by the substitution rules themselves. Whether this is a general property of substitution tilings, or particular to the AB geometry, is the natural next question -- one that the same local tile rearrangements we noted at the outset are well placed to answer, and one that we will address in future work.

\clearpage
\appendix
\onecolumngrid
\renewcommand{\thefigure}{A\arabic{figure}}
\setcounter{figure}{0}
\section{Low energy square supertiles \label{app:square supertiles}}
\begin{figure*}
	\centering
	\includegraphics[width=0.9\linewidth]{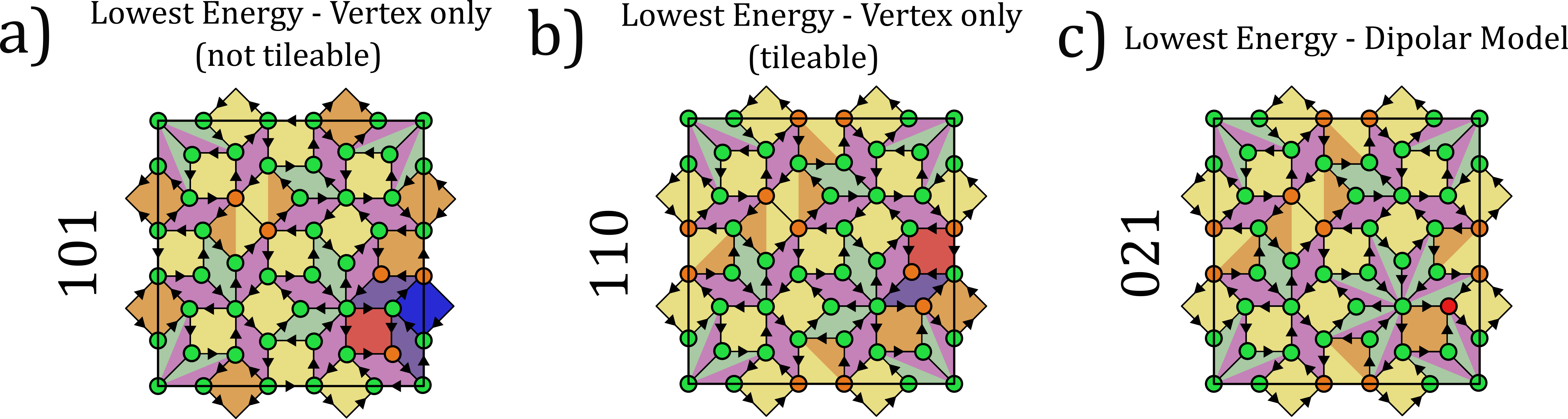}
	\caption{Low energy solutions for the square supertile. (a) The lowest energy configuration when only considering independent vertices. Additional prototiles are included: purple rhomb forming an anti-clockwise loop, blue square forming clock-wise loop, red square with two clockwise and two anti-clockwise spins. The solution cannot tile as the spins won't overlay with the low energy rhomb. (b) The lowest energy configuration for tileable square supertile. (c) The lowest energy configuration when accounting for longer range interactions beyond just the vertex points themselves.}
	\label{fig:sqsupertiles}
\end{figure*}

\begin{figure*}
	\centering
	\includegraphics[width=0.9\linewidth]{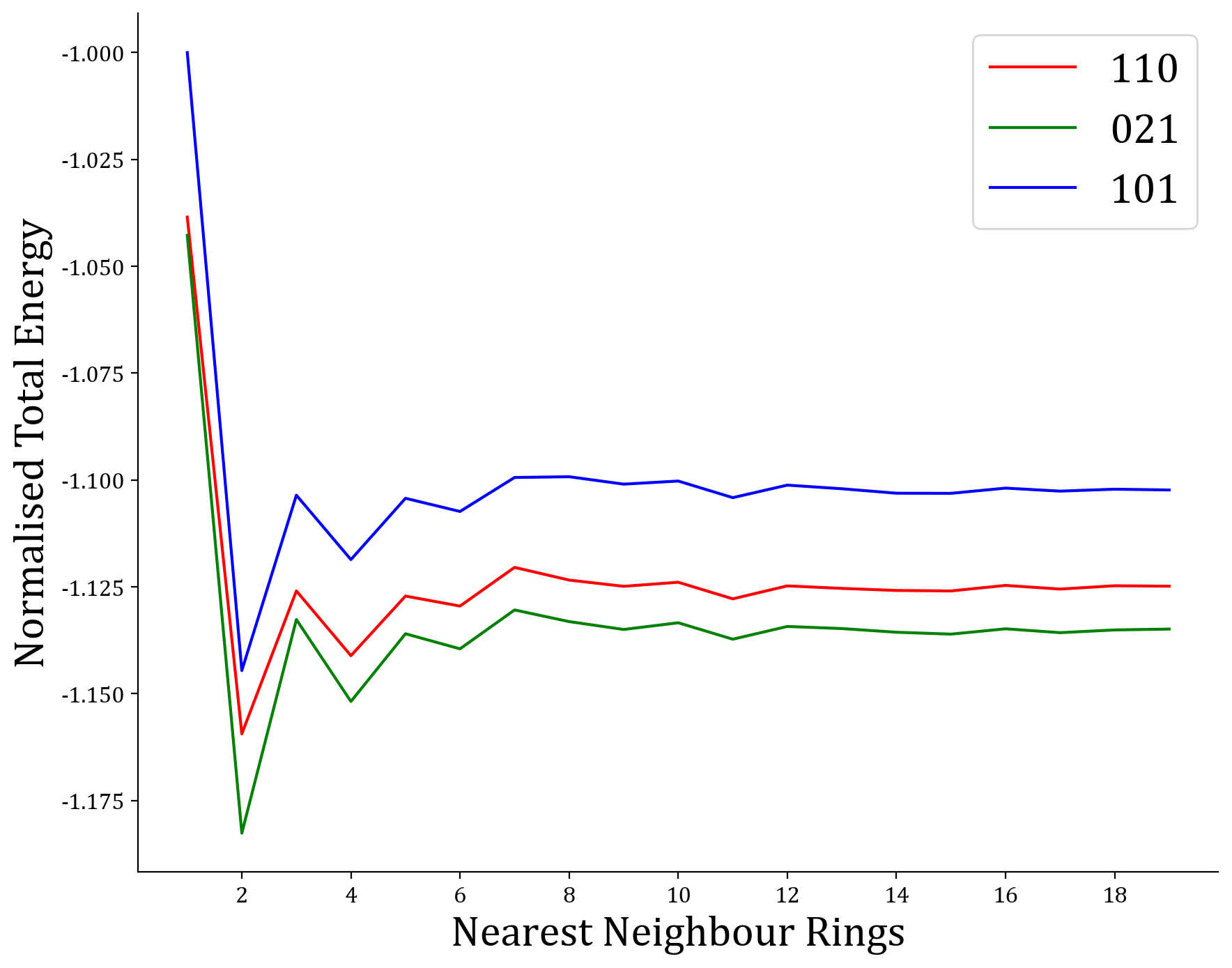}
	\caption{Total energy of each supertile, normalised by the maximum energy. Energy is calculated using the dipolar Hamiltonian, and shown for a varying number of nearest neighbour rings which are included in dipolar coupling. Each ring contains all spins at the same unique distance from each spin, accounting for the varying local environments in the geometry, such that increasing the number of rings increases the number of neighbour spins included in the dipolar coupling. }
	\label{fig:hamil_energy}
\end{figure*}

The simplest route to forming a low energy solution for our ASI geometry is piecing together the low energy forms of each vertex type, as identified with micromagnetic simulations, until topological frustration enforces excitation, in which case we attempt to minimise the energy of such an excitation. Constructing a solution on the substitution rules of a tiling simplifies this approach significantly, as a small cluster of vertices may be decorated to define a solution for the infinite tiling. Applying this approach to the AB ASI, we identify two lowest energy solutions, as shown in Figure \ref{fig:sqsupertiles}(a,b). Figure \ref{fig:sqsupertiles}(a) shows the absolute lowest energy configuration for the square supertile vertices in this way. All but one 5i vertices on the edge are satisfied as E0, with the internal 5i vertices at the E1 level, and two 3i vertices excited to E1. However this decoration cannot tile infinitely, due to the edge decoration. Figure \ref{fig:sqsupertiles}(b) shows the lowest energy configuration with this approach that is able to tile infinitely, with all but one 5i vertices on the edge excited to E1, and a 3i, 4i, and the internal 5i excited to E1. Comparing the total energies of the supertiles in Figure \ref{fig:sqsupertiles}, calculated as a sum of the individual vertex energies, gives that the 101 tile is 0.53eV lower than 110, and 110 is 0.044eV lower than 021. However, the vertex-only approach with which we build these decorations is not physical when considering a long range interacting model. Closely interacting spins, such as those on opposing edges of square prototiles, are not accounted for in this energy minimisation. As such, we must account for longer range interaction effects and propose a new solution more specific to our interaction parameters. This is shown in Figure \ref{fig:sqsupertiles}(b). All 5i in the supertile are excited to E1, with one 4i excited to E2. This decoration notably maximises yellow prototiles over orange prototiles, maximising closed flux loops. Figure \ref{fig:hamil_energy} shows the energy of each supertile, calculated with our dipolar hamiltonian, as we increase the number of nearest neighbour rings included. The 021 supertile is energetically lower for all dopolar distances, with the 101 supertile being the highest.

\clearpage
\newpage 
\bibliography{ref}

\clearpage

\end{document}